\documentclass[12pt,twoside]{article}

\begin{document}

{\bf Locality in Nonsequential Quantum Operations}

\bigskip

\bigskip

{\bf Miroljub Dugi\' c\footnote{Department of Physics, Faculty of
Science, Kragujevac, Serbia and Montenegro, E-mail:
dugic@kg.ac.yu}}

\bigskip

{\bf Abstract:} We give an example of fulfillment of the condition
of locality--no information transfer between certain
subsystems--in a tripartite quantum  system whose dynamics can not
be decomposed (non-sequential dynamics of the system). The three
subsystems ($A$, $B$ and $C$) are designed such that $C$ interacts
simultaneously  with both $A$ and $B$, while there is not any
interaction between $A$ and $B$. On this basis, we emphasize
validity of the condition of locality in a realistic physical
situation.

\bigskip

{\bf KEY WORDS:} Quantum locality; quantum subsystems; global
unitarity

{\bf PACS:} 03.67.-a, 03.65.Ud.

\bigskip

\bigskip

{\bf 1. INTRODUCTION}

\bigskip

\noindent Information transfer is at the heart of information
theory, and particularly the issue of locality--i.e. of dividing
the universe into subsystems for which information transfer may
not be allowed. In a recent article, Schumacher and
Westmorelannd$^{(1)}$ introduced an idea of  {\it dynamical
locality} based on a division of a quantum universe into three
subsystems, and investigate the consequences of the unitary
dynamics of the universe. The condition of locality is described
by the requirement of '{\it no information transfer}' from the
subsystem $B$ to the subsystem $A$ in the composite system
$A+C+B$. They conclude that, under a unitary dynamics for
$A+C+B$, the condition of locality can be satisfied only if
interaction between $A$ and $C$ precedes interaction between $C$
and $B$, while no interaction between $A$ and $B$ is allowed in
the system. Nevertheless, the converse does not hold true.
Actually, for a unitary dynamics of the composite system, the
requirement of locality does not imply sequential interaction as
described above, while this may be obtained for a special initial
state of $C$ (a standard initial state denoted $\vert 0 \rangle
$). This notion exhibits how subtle may be the dynamics of a
tripartite quantum system.

In this paper, we slightly relax and vary some assumptions of the
analysis in Ref. 1, and provide a scenario in which {\it
non-sequential} interaction in a tripartite system may still
allow the 'no information transfer' between the "remote"
(noninteracting) subsystems ($A$ and $B$). More precisely, we
assume the unitary dynamics in the composite system $A+C+B$,
where $A$ and $B$ do {\it not} mutually interact, while $C$ {\it
simultaneously} interacts with both $A$ and $B$. We relax the
condition of arbitrary initial state of $C$ by assuming some
special (cf. below) initial state $\vert 0\rangle$, while
assuming  the interaction between $C$ and $B$ to dominate in the
system. As a result, we obtain that, for some time interval,
there is not any information transfer between $A$ and $B$ in {\it
any direction} possible, while in the limit of the infinite time,
there is a possibility of two-directional information transfer
between $A$ and $B$. This way the condition of 'no information
transfer' may be fulfilled even in case of non-sequential,
non-decomposable interaction in the composite system, and may
even be extended in both directions. We emphasize subtlety of a
tripartite-quantum-system dynamics by illustrating applicability
of our scenario to a physical situation typical for certain
physical-chemistry situations (e.g. an ion/atom in a solution).

\bigskip

{\bf 2. THE LOCALITY CONDITION}

\bigskip

\noindent Let us put the notions of Ref. 1 more precisely and
introduce the notation.

Schumacher and Westmoreland extend the analysis of Beckman et
al$^{(2)}$ that refers to the condition of locality in the
context of quantum operations, in which one deals with an
isolated bipartite system $A+B$. Actually, they point out
necessity of further "coarse graining" of the isolated system,
thus dealing with a tripartite system $A+C+B$, for which the
condition of dynamical locality is investigated in some
generality.

The notion of locality stems: since the state of $B$ does not
alter the state of $A$, {\it no information transfer} is allowed
from $B$ to $A$. That is, in order to know the state of $A$, only
the state of the system $A+C$ is required to be known--as long as
$A$ and $C$ are in interaction. The condition of dynamical
locality can be introduced in a few, mutually equivalent ways.
E.g., nothing we can do to $B$ affects $A$ (the "(Locality III)"
condition)$^{(1)}$.

Now, the task is to investigate the general conditions for the
tripartite system's dynamics that allow the locality condition to
be fulfilled. Under the assumption of the unitary dynamics for
$A+C+B$, the locality condition can be satisfied {\it only} if
the following conditions are fulfilled: (A) the interaction
between $A$ and $C$ precedes interaction between $C$ and $B$, and
(B) $A$ and $B$ are not in mutual interaction. Then, the overall
unitary dynamics can be presented as:

\begin{equation}
\hat U_{ACB} = \left(\hat I_A \otimes \hat W_{CB}\right) \left(
\hat V_{AB} \otimes \hat I_{B}\right)
\end{equation}

\noindent where both $\hat W$ and $\hat V$ represent the unitary
operators of evolution in time; that is, the dynamics is both
decomposable and sequential.

On the other hand, one may wonder: if, from the outset it is
required the locality to be fulfilled, what can be told about the
system dynamics? Interestingly enough, requiring locality to be
fulfilled does {\it not} imply  sequential dynamics as
distinguished in (1). To this end, the different scenarios are
possible, such as$^{(1)}$ a {\it prior interaction} (e.g.
measurement) between $A$ and $B$ (measurement on $A$ performed by
$B$). In summary: the sequential dynamics (1) guarantees validity
of the condition of locality, while the converse does not hold
true.

Nevertheless, despite generality of the analysis$^{(1)}$, there
might intuitively seem many rooms to be left for certain effects
in the tripartite systems not yet obvious or directly predicted by
the general analysis in Ref. 1. And this is the subject of the
next section, in which we give an example of {\it simultaneous}
(and therefore neither decomoposable nor sequential) interaction
between $A$ and $C$, and $C$ and $B$, yet without any prior or
posterior interaction between $A$ and $B$--while the locality
condition is satisfied.

\bigskip

{\bf 3. DISD MODEL}

\bigskip

\noindent We consider a tripartite system $A+C+B$ and we slightly
departure from the general discussion in Ref. 1 as follows: (i)
we assume a special initial state of $C$, and (ii) we assume that
the interaction between $C$ and $B$ dominates in the system.

Let us put these notions in a mathematical form.

First, the {\it initial} state of the composite system is assumed
to read:

\begin{equation}
\vert \Psi \rangle_{ACB} = \sum_i \alpha_i \vert i \rangle_A
\otimes \vert 0 \rangle_C  \otimes \vert \chi\rangle_B
\end{equation}

\noindent bearing obvious notation; $\sum_i \vert \alpha_i
\vert^2 = 1$. The system's Hamiltonian is defined as:

\begin{equation}
\hat H = \hat H_A + \hat H_C + \hat H_B + \hat H_{AC} + \hat
H_{CB}
\end{equation}

\noindent where we assume $\hat H_{AB} = 0$. The crucial
assumptions of our model are as follows: (a) $\hat H_{CB}$
dominates in the system, being described by the coupling constant
$\mathcal{C}_1$, and (b) the initial state $\vert 0 \rangle_C$ is
{\it robust}, relative to the interaction $\hat H_{CB}$--i.e. the
state $\vert 0 \rangle_C$ can not be changed by the interaction
$\hat H_{CB}$: $_C\langle j \vert \hat H_{CB} \vert 0 \rangle_C =
0$, $\forall{j} $ such that $_C\langle j \vert 0 \rangle_C = 0$.

Then, with the aid of the standard (time independent)
perturbation theory$^{(3)}$, it can be shown$^{(4, 5)}$ that the
{\it exact} form of state of the composite system reads:

$$\vert \Psi (t) \rangle_{ACB} = \sum_i \alpha_i (t) \exp (- \imath
t \lambda^{(1)}_{i0} / \hbar) \vert i \rangle_A \otimes \exp (-
\imath t \lambda_{\circ}/ \hbar) \vert 0 \rangle_C \otimes$$
$$\otimes \sum_j \beta_j (t) \exp (- \imath t \lambda_{i0j} /
\hbar ) \vert j \rangle_B + \vert O(\epsilon) \rangle_{ACB}.
\eqno (4)$$

\noindent In eq. (4): $\lambda^{(1)}_{i0} \equiv _{AC}\langle i 0
\vert \hat H_{CB} \vert i 0\rangle_{AC}$ is a part of the first
order correction to the eigenvalues of $\hat H_{CB}$, while
$\lambda_{i0j}$ involves the higher order corrections to the
eigenvalues of $\hat H_{CB}$, and $\lambda_0 =  _C\langle 0\vert
\hat H_C \vert 0 \rangle_C$. The quantity $\epsilon$ (that is
implicit in both $\vert O(\epsilon)\rangle_{ACB}$ and
$\lambda_{i0j}$) is the maximum (or supreme) of the exact
corrections to the eigenstates of $\hat H_{CB}$, its magnitude
being proportional to $\mathcal{C}_2 / \mathcal{C}_1$:
$\mathcal{C}_2$ and $\mathcal{C}_1$ are the coupling constants of
$\hat H_{AC}$ and $\hat H_{CB}$, respectively. The constants are
defined as follows: $\alpha_i (t) \equiv \alpha_i \exp (- \imath
t _A\langle i\vert \hat H_A \vert i \rangle_A)$, while $\beta_j
(t) \equiv \beta_j \exp [- \imath t (E^{(0)}_{0j} + _B\langle
j\vert \hat H_E\vert j\rangle_B)/ \hbar ]$; $E^{(0)}_{0j}$ is an
eigenvalue of $\hat H_{CB}$, while $\sum_j \vert \beta_j \vert^2
= 1$. [For more details cf. Refs. 4 and 6.]

As it can be shown$^{(4)}$, $\Vert \vert O(\epsilon) \rangle \Vert
\sim \mathcal{C}_1^{-1}$ as well as $\lambda \sim
\mathcal{C}_1^{-1}$, where $\lambda \equiv sup \{\lambda_{i0j}\}$.
Then, if $\mathcal{C}_2 / \mathcal{C}_1 \to 0$, one may state
$\Vert \vert O(\epsilon) \rangle \Vert \to 0$ and $\lambda \to
0$, which gives {\it exactly} rise to the lack of correlations in
the composite system. For the realistic situations, i.e. when
$\mathcal{C}_2 / \mathcal{C}_1 \ll 1$, while neglecting the small
term in (4), one may state the lack of correlations in the time
interval $\tau$ less or of the order of $\lambda^{-1}$. For the
arbitrarily long time interval much longer than $\tau$, there
appear the correlations in the composite system. Needless to say,
given (4), it is almost trivial to prove the told within the
quantum operations formalism (that assumes the use of the "trace"
operation).

Physically, for sufficiently strong interaction $\hat H_{CB}$, the
system $A$ evolves  in time approximately {\it
unitarily-like}--as if it were not being an open system--for the
period of time of the order of $\tau$. And this is the central
observation of the model presented.

Here, we want to emphasize: the choice of the initial state as
given in (2) need not be physically artificial. First, this choice
is adapted to the general analysis$^{(1)}$ we start from. Second,
the special state $\vert 0 \rangle_C$ is in accordance with the
foundations of the decoherence theory$^{(7)}$: the "environment"
$B$ may select such a special state for the (open) system $C$, as
discussed in Refs. 4 and 6. Therefore, the model (the so-called
$^{(4, 5)}$ DISD model) discussed here seems applicable to the
realistic physical situations as discussed briefly in Section 4.
Third, the initial state of $C$ may externally and locally be
prepared.

The lack of correlations between $A$ and $B$ clearly stems the
condition of locality: e.g. reading out a state of $B$ (of $A$)
does not provide any information about the state of $A$ (of
$B$)--the "(Locality I)" condition$^{(1)}$. This is even stronger
condition of locality than the one investigated in Ref. 1: there
is not any information transfer between $A$ and $B$, in any
direction possible. Interestingly enough, the interaction in the
system is neither decomposable nor sequential. Certainly, for the
arbitrarily long time interval (much longer than $\tau$ defined
above), the system $C$ mediates the information transfer: despite
the fact that $A$ and $B$ do not mutually interact, the system $C$
mediates the interaction and provides the correlations of states
between $A$ and $B$ --cf. the factor $\lambda_{i0j}$ in (4).
Depending on the details in the model (the kind of interactions,
the coupling constants and their ratio, the initial states of the
subsystems etc.), there might appear variety of the different
effects in the system. Our analysis distinguishes such an effect
of interest: the dynamics of $A+C+B$ system can {\it not} be
decomposed, while keeping $A$ and $B$ dynamically (and
information-theoretically) separated, where interaction between
$A$ and $B$ is {\it switched off} in the infinite time, $t \in (-
\infty, \infty)$, as long as the initial state (2) can be
independently--i.e. without interaction between $A$ and
$B$--prepared.

\bigskip

{\bf 4. DISCUSSION AND CONCLUSION}

\bigskip

\noindent What is the place of our model in the context of the
general discussion in Ref. 1?

We use the condition of locality as {\it additional requirement}
to the unitary dynamics of the composite system. As mentioned
above (cf. Section 2), then the sequential interaction in the
composite system is not required$^{(1)}$. And this is exactly the
point at which our considerations fit the general discussion
presented in Section 2. Actually, in this context, we relax
certain assumptions of the general discussion and it is therefore
not for surprise that we obtain a hopefully interesting result,
still extending the notion of locality: not only $A$ is 'local a
system' relative to $B$, but the converse holds true. The
conditions we have in mind are the points (i) and (ii) in Section
3. Our result reads: for some time interval\footnote{In the limit
$\mathcal{C}_2 / \mathcal{C}_1 \to 0$, this time interval becomes
infinitely long.}, the system $A$ behaves effectively as if it
were an isolated system, thus not providing any information for
$B$, neither $B$ may provide information transfer to $A$. For
arbitrarily long time interval, however, there inevitably appear
the correlations between $A$ and $B$, thus providing in principle
the two-directional information transfer: e.g. reading out the
state of $A$ (of $B$) one may conclude about the state of $B$ (of
$A$).

Physically, the situation described in Section 3 directly refers
to the following issues. Originally, the
DISD\footnote{Originating from the "decoherence-induced
suppression of decoherence"$^{(4)}$.} method was developed for
the purposes of combating decoherence in the quantum computers
hardware$^{(4)}$-- a short version may be found in Ref. 5.
Nevertheless, the model bears generality, referring to both,
suppressing the quantum entanglement in a tripartite quantum
system$^{(6)}$ as well as to the issue of {\it avoiding
decoherence}--the issue of the decoherence control. As to the
former, the model is directly applicable to the following
physical situation$^{(6)}$: an atom (or ion--the system $A$) is
surrounded by a cage of the solvent molecules (system $C$),
which, in turn is in strong interaction with the rest of the
solution (the cage's environment--the system $B$). Assuming that
the system $C$ is sufficiently macroscopic--as the general
decoherence theory$^{(7)}$ stems--the system $B$ may be able to
select\footnote{This situation is mentioned in Section 2: there
is a prior interaction between $A$ and $B$: then, the time
interval of interest is $t \in [0, \infty)$, which challenges
validity of (2).} a special state $\vert 0 \rangle$ of $C$--i.e.
$B$ may decohere $C$ (which is the origin of the acronym DISD--cf.
footnote 3). Due to the strong interaction between $C$ and $B$,
this state of $C$ may remain intact ("robust")$^{(8)}$, i.e.
unchanged in time, thus guaranteeing$^{(4-6)}$ the unitary-like
dynamics for $A$. Needless to say, then $A$ remains effectively
decoupled from $B$ for a time interval of the order of $\tau$
(cf. above). To this end, the individuality of the atom as well as
non-transfer of information (from $A$ to $B$ and {\it vice versa},
and virtually to $C$) represents an example of satisfiability of
the locality condition in a realistic physical situation.

Therefore, we conclude, that subtlety of quantum dynamics of the
tripartite systems leaves much room for a variety of dynamical
effects in the real systems. As we show, slightly relaxing the
general assumptions of the dynamical locality model of Schumacher
and Westmoreland$^{(1)}$, we are able to obtain the dynamical
effects not directly predicted within the context of the general
discussion. Actually, we provide an example for non-necessity of
the sequential interaction {\it while} the condition of (extended)
dynamical locality is satisfied, as well as the possibility of
two-directional information flow mediated by the system $C$ (for
arbitrarily long time interval), both within the context of the
{\it same model} of a tripartite system $A+C+B$.

\bigskip

{\bf ACKNOWLEDGEMENTS}

\bigskip

\noindent The work on this paper was supported by Ministry of
Science and Environmental protection, Serbia.

\bigskip

{\bf REFERENCES}

\bigskip

1. B. Schumacher and M. D. Westmoreland, {\it Quantum Information
Processing} {\bf 4}(1), 13 (2005)

2. D. Beckman, D. Gottesman, M. A. Nielsen, and John Preskill,
{\it Phys. Rev. A} {\bf 64}, 052309 (2001)

3. A. Messiah, {\it Quantum Mechanics} (North Holland Publishing
Company, Amsterdam, 1976)

4. M. Dugi\' c, {\it Quantum Computers \& Computing} {\bf 1}, 102
(2000)

5. M. Dugi\' c, e-print arXive quant-ph/0001009

6. M. Dugi\' c, {\it Europhys. Lett.} {\bf 60}, 7 (2002)

7. D. Giulini, E. Joos, C. Kiefer, J. Kupsch, I.-O. Stamatescu and
H. D. Zeh, {\it Decoherence and the Appearance of a Classical
World in Quantum Theory} (Springer, Berlin, 1996)

8. W. H. Zurek, {\it Prog. Theor. Phys.} {\bf 89}, 281 (1993)

\end{document}